\documentclass[fleqn,10pt]{wlscirep}
\usepackage[utf8]{inputenc}
\usepackage[T1]{fontenc}
\usepackage{bm}
\usepackage{graphicx}
\usepackage{epstopdf}
\usepackage{amssymb}
\usepackage{amsmath}
\usepackage{xspace}
\usepackage{amsfonts}
\usepackage{mathrsfs}
\usepackage{color}
\usepackage[normalem]{ulem}
\usepackage{phfqit}
\usepackage[utf8]{inputenc}
\usepackage{amssymb}

\title{Polariton-based quantum memristors }

\author[1,$\ast$]{Ariel Norambuena}
\author[2,3,4]{Felipe Torres}
\author[4]{Massimiliano Di Ventra}
\author[1,$\dagger$]{Ra\'ul Coto}
\affil[1]{Centro de Investigaci\'on DAiTA Lab, Facultad de Estudios Interdisciplinarios, Universidad Mayor, Chile}
\affil[2]{Centro de Nanociencia y Nanotecnolog\'ia CEDENNA, Avda. Ecuador 3493, Estaci\'on Central, 9170124, Santiago, Chile}
\affil[3]{Departamento de F\'isica, Facultad de Ciencias, Universidad de Chile, Casilla 653, Santiago, 7800024, Chile}
\affil[4]{Department of Physics, University of California, San Diego, La Jolla, CA 92093,
USA}

\affil[$\ast$]{e-mail: ariel.norambuena@umayor.cl}
\affil[$\dagger$]{e-mail: raul.coto@umayor.cl}

\begin{abstract}
Information processing and storing by the same physical system is emerging as a promising alternative to traditional computing platforms. In turn, this requires the realization of elementary units whose memory content can be easily tuned and controlled. Here, we introduce a polariton-based quantum memristor where the memristive nature arises from the inter-cavity polariton exchange and is controlled by a time-varying atom-cavity detuning. A dynamical hysteresis is characterized by the fluctuations in the instantaneous polariton number, where the history information is encoded into a dynamical phase. Using a Lindblad master equation approach, we find that features of the quantum memristor dynamics, such as the area and circulation of the hysteresis loop, showcase a kind of ``plasticity'' controlled by quantum state initialization. This makes this quantum memristor very versatile for a wide range of applications. 
\end{abstract}
\begin{document}

\flushbottom
\maketitle

\thispagestyle{empty}

\section*{Introduction}

Computing in and with memory~\cite{diventra13a} is emerging as an alternative to our traditional model of computation, which separates the tasks of storage and processing of information into two physically distinct units. Memory (time non-locality) can be realized in many ways and with different types of systems and devices (see, e.g., Refs.~\cite{11_memory_materials} for a review). However, so far particular attention has been mainly given to {\it classical} resistive memories (sometimes called ``memristive elements'') as components of processors with memory~\cite{UMM}. On the other hand, quantum dynamics may offer additional benefits for the realization of new storage and information processing capabilities with applications in quantum information and (neuromorphic) quantum computing~\cite{Pfeiffer, Gonzalez, Salmilehto, Grollier}. 

In this respect, initial suggestions of quantum memristive elements have ranged from superconducting quantum circuits~\cite{Peotta,Pfeiffer, Gonzalez, Salmilehto} to quantum photonic devices~\cite{Sanz}. In all these cases, the memory mechanism arises from the combination of quantum feedback~\cite{Pfeiffer, Salmilehto} and dissipative effects~\cite{Sanz}. However, one can also rely on quasi-particle dynamics, such as low-energy quasi-particle tunneling~\cite{Salmilehto} and exciton-polariton interactions~\cite{Mirek}, as other types of quantum features that can be harnessed to realize quantum memristors. \par

Among quasi-particles, those originating from strongly correlated light-matter systems, known as polaritons, offer attractive technological features like room temperature operation, high dynamical speed, and suitable implementation \cite{Boulier,Basov}. Polaritons have been implemented in a wide variety of systems, including atoms~\cite{Birnbaum}, excitons~\cite{Weisbuch}, trapped ions~\cite{Toyoda} and superconducting circuits~\cite{Raftery}, and have allowed the observation of new quantum phases~\cite{Hartmann_2006,Greentree,Fitzpatrick,Kasprzak_2006}. \par 

Recently, more versatile polariton systems have been designed by considering driving schemes~\cite{Johansen,Clark,Mundada}. The external driving sets a different time scale. Thereby, the system's time evolution is governed by the difference between the intrinsic response time scale of the system and the driving time scale, leading to bistability. Bistability is a hallmark of driven non-linear classical and quantum systems~\cite{Gibbs, Felber, Lugiato, Rempe, Gripp, Rodriguez}. In the latter, this is due to the presence of two stable branches where quantum fluctuations switch the states of the system between these two stable solutions~\cite{Rodriguez, Huybrechts}. Furthermore, the endeavors to address the response of a system under the influence of an external driving focusing on understanding the out-of-equilibrium dynamics have opened new perspectives. In such direction, we found the control of Mott-insulator and superfluid states~\cite{Tancara} and the appearance of dynamical hysteresis in photonic devices~\cite{Rodriguez,Casteels}. 

In this work, we introduce a polariton-based quantum memristor (PQM) whose behavior is controlled by a time-dependent atomic modulation providing a hysteretic response over the variance of the number of quasi-particles. The underlying mechanism relies on the inter-cavity polariton exchange (hopping) and atomic modulation. We demonstrate that memory effects in our PQM depend on the initialization of the system (plasticity) and the time scale of the driving. This minimalistic setup exploits the inter-cavity polariton exchange, leading to interesting phenomena in more complex optical arrays.

\begin{figure}[ht]
\centerline{\includegraphics[width=0.6\textwidth]{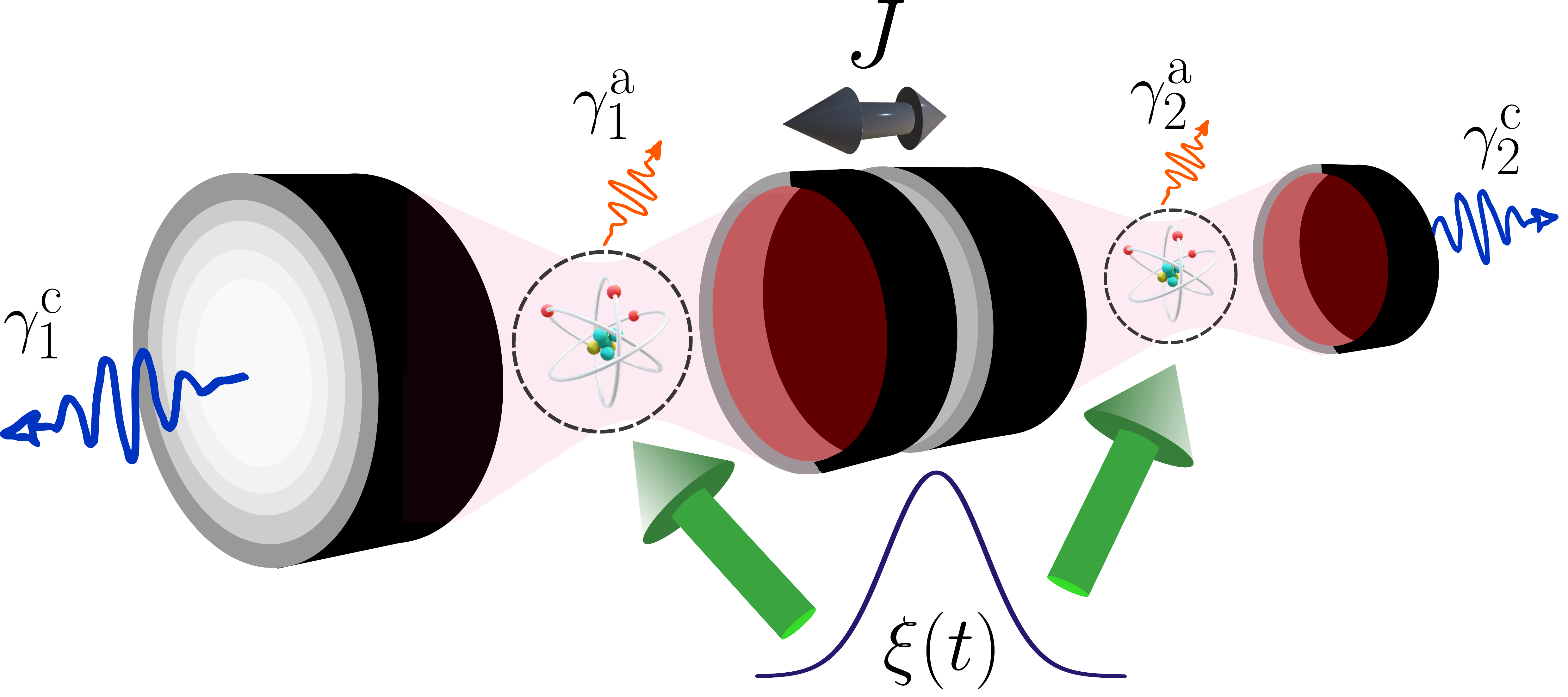}}
\caption{Schematic representation of the polariton-based quantum memristor composed of two coupled cavities QED with atomic modulation $\xi(t)$. Each cavity interacts with a two-level atom, and photon hopping is allowed between the cavities with coupling strength $J$. Furthermore, each cavity experiences atomic $\gamma_i^{\rm a}$ and photonic $\gamma_i^{\rm c}$ losses.} 
\label{fig_cav}
\end{figure}

\section*{Polariton-based quantum memristor}
The envisioned structure consists of two interacting optical cavities driven by a time-dependent atomic modulation $\xi(t)$. Figure~\ref{fig_cav} shows a schematic representation of the PQM composed of two cavities doped with a two-level system (we label $\varphi_g$ and $\varphi_e$), where the intra-cavity light-matter coupling allows the generation of cavity QED polaritons quasi-particles, and the photon hopping between cavities leads to polariton exchange. Dissipative effects arise from imperfect mirrors with photonic decay $(\gamma_i^{\rm c})$ and spontaneous emission with atomic decay $(\gamma_i^{\rm a})$ per site ($i=1,2$). The modulation serves different purposes such as manipulation of qubits~\cite{Tancara,Strand}, sideband transitions~\cite{Beaudoin} and control of atomic quantum gases~\cite{Eckardt}. The dynamics of the light-matter interaction is given by the Jaynes-Cummings-Hubbard (JCH) Hamiltonian, 

\begin{equation}\label{Hamiltonian}
H_0 = \sum_{i=1}^{2}\left[\omega^{\rm c} a_i^{\dagger}a_i + \omega^{\rm a}
\sigma_{i}^{\dagger}\sigma_{i} + g\left(a_i \sigma_{i}^{\dagger} + a_i^{\dagger}\sigma_i \right) \right]- J \left(a_1 a_2^{\dagger} + a_2^{\dagger} a_1\right),
\end{equation}
here $\sigma_{i} = |\varphi_e\rangle_i\langle \varphi_g|_i$ are Pauli operators for the two-level atom at the $i$-th cavity,  $a_i^{\dagger}$ and $a_i$ are photon creation and annihilation operators, respectively, $\omega^{\rm c}$ ($\omega^{\rm a}$) is the cavity (atom) frequency, $g$ is the atom-field coupling strength, and $J$ is the photon hopping strength between neighboring cavities. In this work, we set $\omega^{\rm c}=\omega^{\rm a}$ unless stated otherwise. The total Hamiltonian of the system is modelled as
\begin{equation}
H(t)= H_0+\xi(t)\sum^{2}_{i=1}\sigma_{i}^{\dagger}\sigma_{i}.
\label{eq:1}
\end{equation}

We investigate the dynamics induced by a Gaussian modulation given by

\begin{equation}\label{detuning}
\xi(t) = \xi_i + (\xi_f-\xi_i)e^{-(t-T)^2/2\sigma_w^2},
\end{equation}
where $\xi_f>\xi_i >0$ and $\sigma_w$ is the characteristic width of the Gaussian modulation with the property. We explore the dynamics in the regime $T^2/(2\sigma_w^2)>1$ leading to a Gaussian profile satisfying $\xi(0) = \xi(2T) \approx \xi_i$. We introduce the instantaneous polaritonic basis $|n+\rangle = \sin \theta_n(t)|n,\varphi_g\rangle + \cos \theta_n(t)|n-1,\varphi_e\rangle$ and $|n-\rangle = \cos \theta_n(t)|n,\varphi_g\rangle - \sin \theta_n(t)|n-1,\varphi_e\rangle$ with $\theta_n(t) = (1/2)\mbox{tan}^{-1}(2\sqrt{n}g/\xi(t))$, and instantaneous eigenenergies $E_{n\pm}(t) = \omega^{\rm c}n+\xi(t)/2 \pm (\xi^2(t)+4g^2n)^{1/2}/2 $, where $n$ is the number of photons per cavity. Moreover, sudden inter-branch transitions could be activated if $\xi(t)$ has some non-adiabatic features (see Supplementary Information for further details).

\begin{figure}[ht]
\centerline{\includegraphics[width=0.8\textwidth]{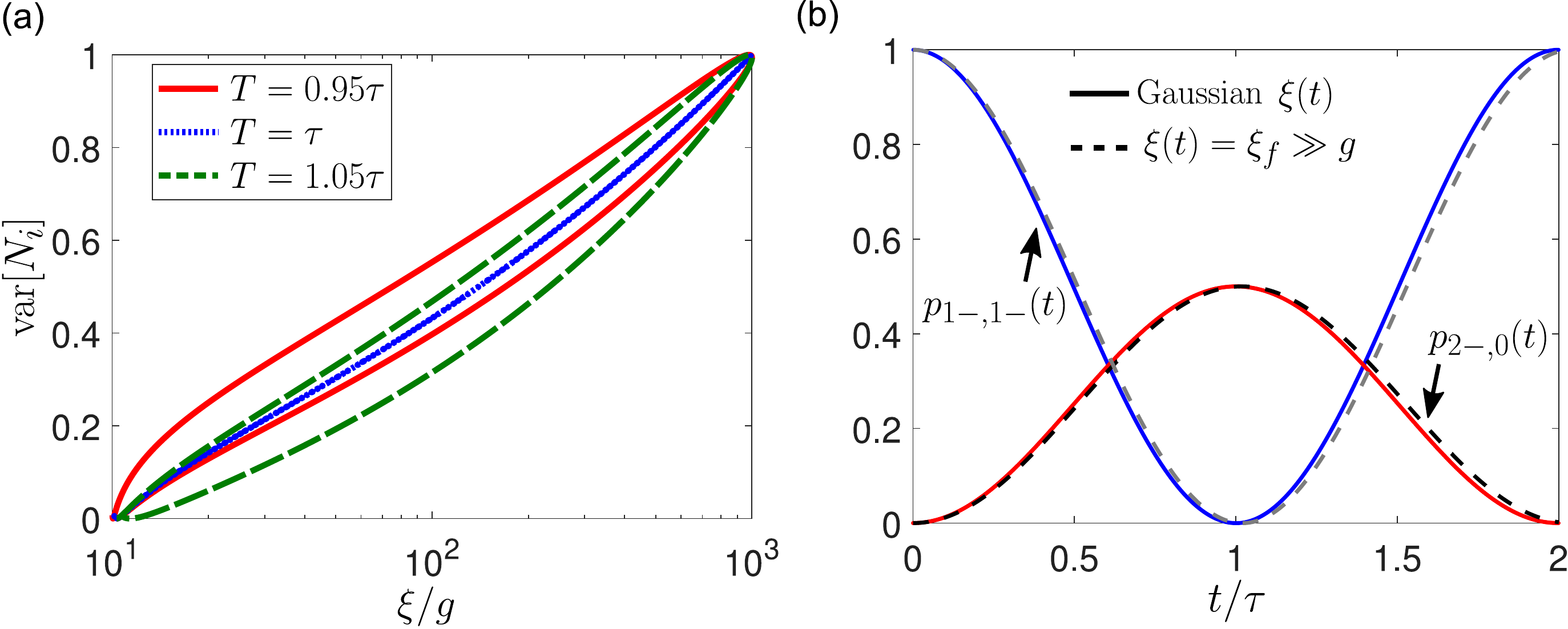}}
\caption{(a) Dynamical hysteresis curves for a closed PQM with $g=1$, $J=10^{-2}g$, $\xi_i = 10g$, $\xi_f = 10^{3}g$, $\sigma_w = T/4$, $\omega^{\rm c} = 10^4g$, and three different times $T$ in the Gaussian modulation. The initial state is a Mott-insulator-like state $|\Psi(0)\rangle = |1-\rangle \otimes |1-\rangle$ for $\xi(t=0) = \xi_i$. (b) Population dynamics. Solid and dashed lines correspond to the exact numerical solutions for the Gaussian time-dependent and a large constant detunings, respectively. The case $\xi(t) = \xi_f \gg g$ shows differences with the time-dependent case, since dynamical phases introduce a small shift in the populations.} \label{fig_hys}
\end{figure}

When the atomic and photonic frequencies are similar this system exhibit a Mott-phase characterized by localized excitations caused by the photon blockade, i.e., the effective repulsion between the excitations~\cite{Angelakis, Birnbaum}, that prevents fluctuations of the total number of polaritonic excitations $\sum_i N_i$ with $N_i = a_i^{\dagger} a_i + \sigma_i^{\dagger} \sigma_i$ the number of quantum excitations in the $i$th cavity. A photonic superfluid phase, characterized by delocalized excitations, emerges from the detuning between atomic and photonic frequencies resembling the flow of charge in a classical resistive memory. During the Mott-superfluid phase transition, the delocalization of the excitations and the polariton exchange lead to an effective quasi-particle flow throughout the optical cavities. In our system, time non-locality arises from the atomic modulation, and memory effects are studied using the variance in the polariton number $\text{var}[N_i(t)] = \mbox{Tr}\left[N_i^2 \rho(t) \right] - \mbox{Tr}\left[N_i\rho(t) \right]^2$.  \par

We describe the open dynamics using the Markovian master equation $\dot{\rho} = -i[H(t),\rho] + \mathcal{D} \rho$, where losses are modeled using the Lindblad form:
\begin{eqnarray} \label{Losses}
{\cal D}\rho &=&  \sum^2_{i=1} \left[\gamma_i^{\rm c} a_{i} \rho a_i^{\dagger} +
\gamma_i^{\rm a} \sigma_{i} \rho \sigma_i^{\dagger}- \frac{1}{2}\left\{a_i^{\dagger} a_i+\sigma_i^{\dagger} \sigma_i  ,\rho \right\}  \right].\nonumber\\
&&
\label{dissipator}
\end{eqnarray}
For concreteness, we first consider an initial Mott-insulator-like state $|\Psi(0)\rangle = |1-\rangle^{\xi(0)} \otimes |1-\rangle^{\xi(0)}$, where $|1-\rangle^{\xi(0)}$ means that the initial state is calculated for $\xi(0)$. Additionally, we set $\xi(t)\geq 10g$ in Eq.~\eqref{detuning}, which serves two purposes. Firstly, it enables inter-cavity polariton exchange by suppressing the photon-blockade. Secondly, variations on $\theta_n(t)$ are small, constraining the effect of the modulation to small shifts on the variance. In Fig.~\ref{fig_hys}(a) we show that the evolution of $\text{var}[N_i(t)]$ describes a hysteresis loop as a function of $\xi(t)$. We remark that the transition Mott-Superfluid takes place regardless the instantaneous value of $\xi(t)$. However, the driving time scale originates a shift on the population, which can be observed in Fig~\ref{fig_hys}(b) and Eqs.(S11)-(S12) in the Supplementary Information. Therefore, as the position of the Gaussian peak ($T$) changes different loops appear, but there is a critical value ($\tau = \pi/(4J) \approx 78.5 g^{-1}$ for $J = 10^{-2}g$) that suppress the hysteretic behavior (see Supplementary Information for the estimation of the critical time). To understand this, we provide a simple argument based on the time-symmetry of the functions involved.  \par

Let us consider the time evolution of the occupation probabilities depicted in Fig.~\ref{fig_hys}(b). Throughout the manuscript, we refer to the occupation probability for each instantaneous two-body polaritonic state $|n \pm, m\pm\rangle$ as $p_{n\pm, m\pm } = \langle n \pm,m\pm| \rho(t)|n\pm,m \pm\rangle$. It is worth noticing that there is a characteristic time ($\tau$) that corresponds to the condition $p_{1-,1-}(\tau)=0$, where the system is in a superfluid-like state, \textit{i.e.} only states $|2-,0g\rangle$ and $|2-,0g\rangle$ (with the same probabilities) are present. Around this time ($\tau$) the dynamics is roughly time-symmetric, $\text{var}[N_i(\tau+t)] \approx \text{var}[N_i(\tau-t)]$. Moreover, the detuning in Eq.~\eqref{detuning} is a Gaussian function, and it is symmetric around $T$. Hence, when $T=\tau$, $\text{var}[N_i(t)]$ and $\xi(t)$ exhibit the same time-symmetry which makes the hysteresis loop vanish, see Fig.~\ref{fig_hys}(a). However, as $\tau$ and $T$ differ, the hysteresis loop is restored. The hysteresis loop arises from the delay between the driving time ($T$) and the time-scale of the system ($\tau$). This implies that the irreversibility of the dynamical process is produced by the response time scale of the system against external driving. We analyze the hysteresis loop area and its circulation for different initial conditions in the closed and open dynamics. In the Supplementary Information we analyze the characteristic time $\tau$ in an effective spin-$1$ system that mimics our PQM. \par

\begin{figure}[h!]
\centerline{\includegraphics[width=0.75 \linewidth]{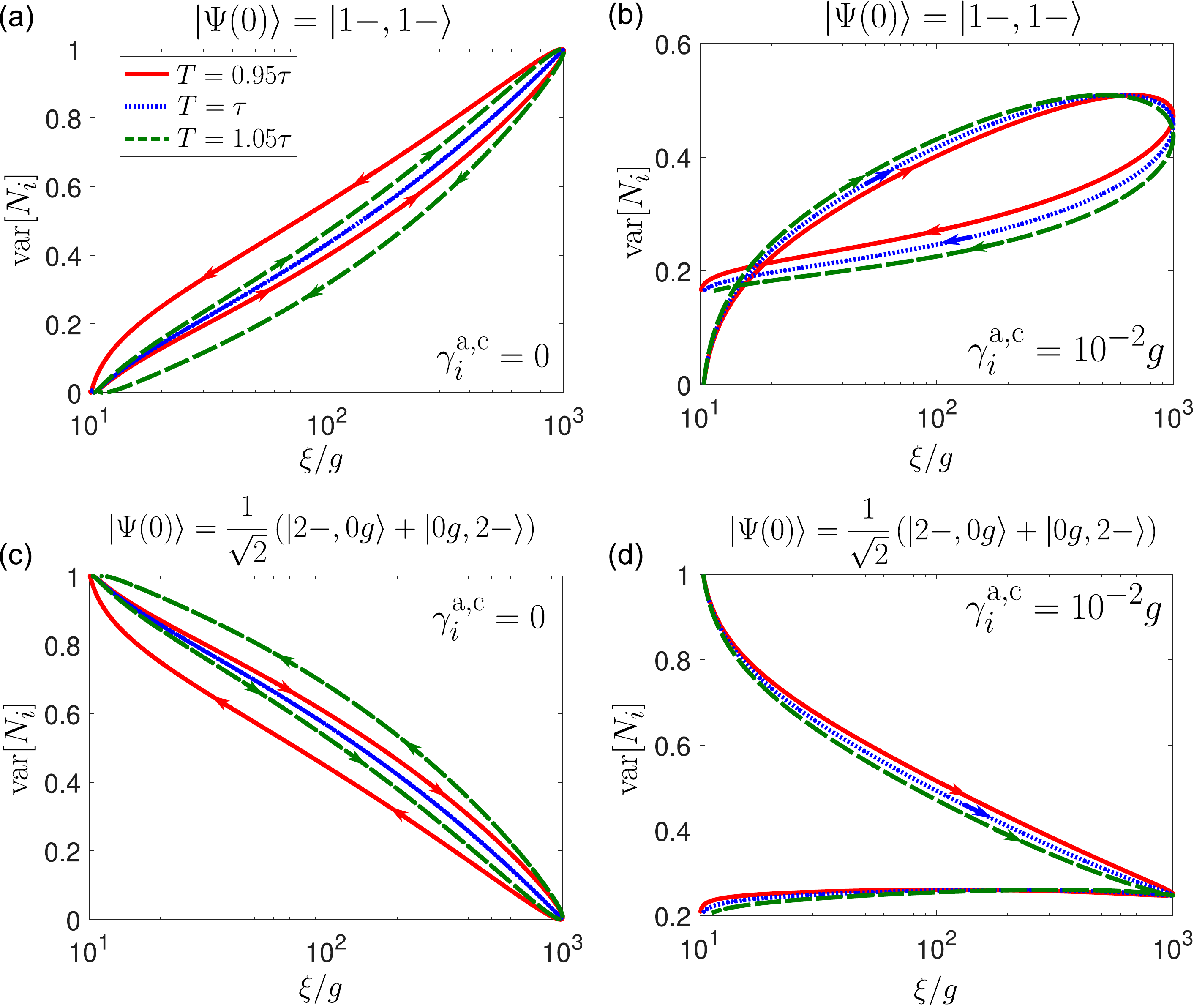}}
\caption{Hysteresis curves for the Mott insulator (superfluid) initial condition $|\Psi(0)\rangle = |1-,1-\rangle$ ($|\Psi(0)\rangle = (1/\sqrt{2})[|2-,0g\rangle+|0g,2-\rangle]$) for the closed ((a) and (c)) and open ((b) and (d)) dynamics of the PQM. For all simulations the parameters are $g=1$, $J=10^{-2}g$, $\xi_i = 10g$, $\xi_f = 10^{3}g$, $\sigma_w = T/4$, and $\omega^{\rm c} = 10^4g$. For the open case we set $\gamma_i^{\rm a,c} = 10^{-2}g$ for each cavity $i=1,2$.}
\label{fig_hys_open}
\end{figure}

\section*{Hysteresis loop area and circulation}

The oriented area of a simple connected curve $(x(t),y(t))$ can be defined as $-\oint y dx$. In our particular case, we have $x(t) = \xi(t)$ (control) and $y(t) = \mbox{var}[N_i]$ (response). The enclosed area in the absence of dissipation is approximately given by
\begin{equation} \label{Area_exp_final}
A  \approx {\eta \over \sqrt{2}}(\xi_f - \xi_i) \pi^{3/2}{\sigma_w \over \tau}\sin \left( {\pi T \over \tau} \right)e^{-(\pi \sigma_w/(\sqrt{2}\tau))^2},
\end{equation}
where $\eta = 1-2\mbox{var}[N_i(0)]$ is a factor that depends on the state initialization (see Supplementary Information). We first notice that $A=0$ for $T=\tau$ ($\sin(\pi T/\tau) = 0$), which explains why the hysteresis vanishes in Fig.~\ref{fig_hys}(a) dotted line. When $T \neq \tau$, the sign of $A$ accounts for the direction of the circulation, where $A>0$ (anti-clockwise) and $A<0$ (clockwise). We analytically find that the sign of the area depends on three factors, namely, the variation $\xi_f - \xi_i$ (which is positive in our work), the ratio $T/\tau$, and the factor $\eta$ that accounts for state initialization. We show in Fig.~\ref{fig_hys_open} that the inter-cavity polariton exchange and the time-dependent driving $\xi(t)$ induce different responses in the variance of the polariton number $\text{var}[N_i]$ depending on the initial state $\rho(0) = |\Psi(0)\rangle \langle \Psi(0)|$. \par

In Fig.~\ref{fig_hys_open}(a) we show the case where $\eta = 1$ ($\mbox{var}[N_i(0)] = 0$), leading to anti-clockwise (clockwise) circulation for $T<\tau$ ($T>\tau$). In Fig.~\ref{fig_hys_open}(c), instead, we show the case $\eta = -1$ ($\mbox{var}[N_i(0)] = 1$), for which the circulation has been inverted with respect to the previous case. In Fig.~\ref{fig_hys_open}(b)-(d) we show the hysteresis for the open dynamics. The main difference is that there is still a loop in Fig.~\ref{fig_hys_open}(b), while in the Fig.~\ref{fig_hys_open}(d), the loop breaks. The latter is a consequence of the initial condition that sets the initial variance ($\text{var}[N_i(0)]$) at the maximum value, while losses do not allow the system to reach that state again. Finally, we remark that the dynamics is no longer symmetric around $\tau=78.5g^{-1}$, thus the condition $T=\tau$ for shrinking the area fails. \par

\begin{figure}[ht]
\centering
\includegraphics[width=0.75 \linewidth]{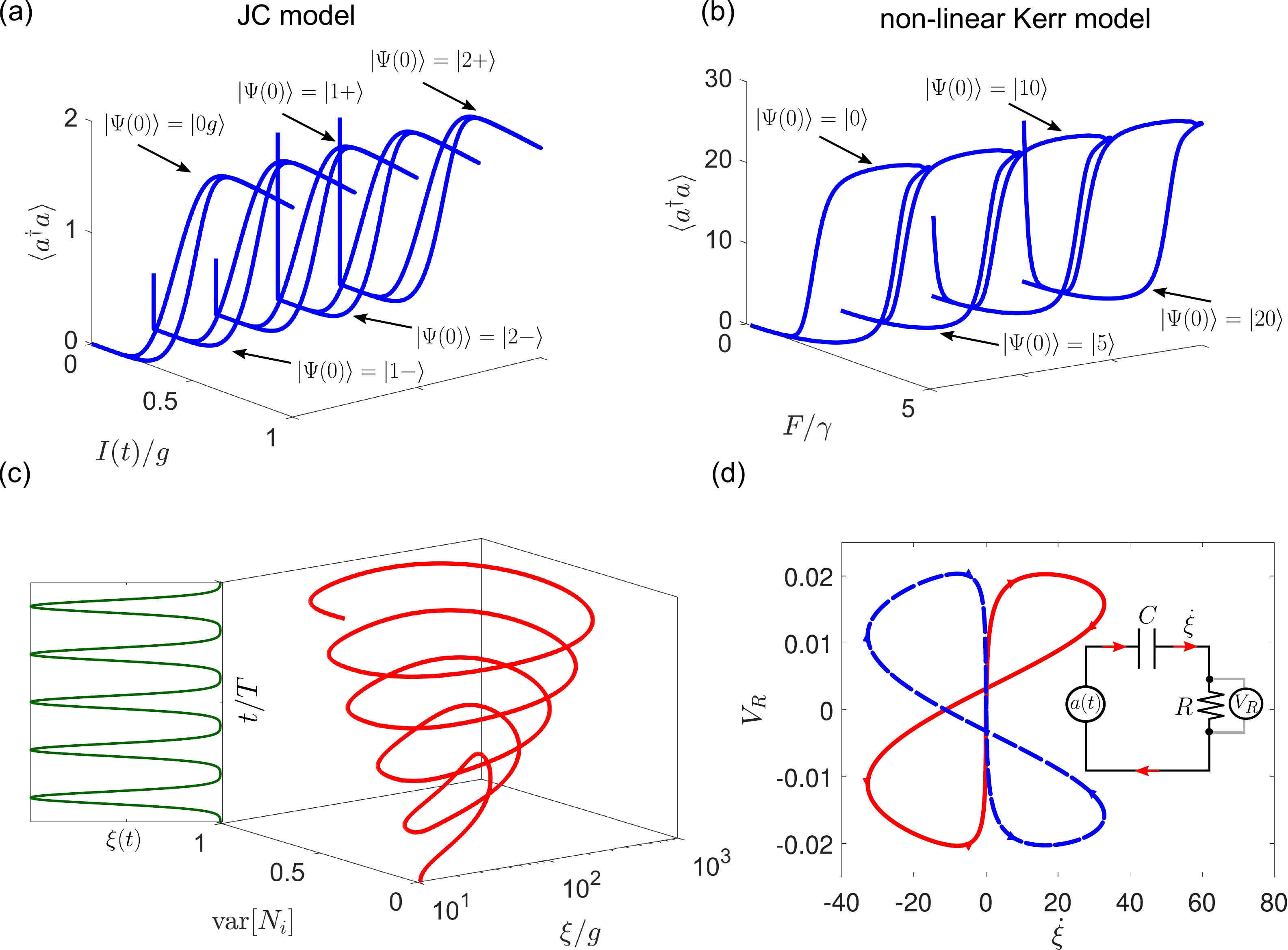}
\caption{(a) Hysteresis curves for the Jaynes-Cummings modulated with a time-dependent laser with frequency $\omega_l$ and a Gaussian laser amplitude $I(t)$. The hysteresis curves are plotted for the mean number of photons $\langle a^{\dagger}a\rangle$ as a function of the laser amplitude $I(t)/g$ using different polaritonic initial conditions $|\Psi(0)\rangle$. For the simulation we use the parameters $g=1$, $w^{\rm a} = w^{\rm c} = w^{\rm l} = 10^4 g$, $\gamma^{\rm a} = \gamma^{\rm c} = g/10$ (photonic and atomic losses) and $T=0.95 \tau$. (b) Hysteresis curves for the non-linear Kerr model introduced in Ref.~\cite{Casteels}  with time-dependent laser amplitude $F(t)$ following a triangular shape with sweep time $t_s = 10\gamma^{-2}\Delta F$, $\Delta F = 3 F_0$, $F_0 = \gamma$. The mean number of photons $\langle a^{\dagger}a\rangle$ as a function of $F/\gamma$ shows a closed hysteresis loop. For the simulations we consider $\gamma = 1$, $\Delta = 2 \gamma$, $U = 0.1\gamma$, $\gamma = 1$ and different initial Fock states $|\Psi(0)\rangle$. (c) Periodic modulation $\xi(t)$ and its effects on the hysteresis curve $(\xi(t)/g,\mbox{var}[N_i])$ as a function of time $t/T$. The 3D plot show how the response function $\mbox{var}[N_i]$ describes a spiral when the train-pulse $\xi(t)$ is introduced. For the simulation we consider a closed dynamics with $g= 1$, $J = 10^{-2}g$, $\omega^{\rm c} = 10^{4}g$, $T = 0.95 \tau$, $\sigma_w = T/4$, and the initial condition $|\Psi(0)\rangle = |1-,1-\rangle$. (d) PQM dynamics simulated from Eq.~(\ref{memristor}) starting from the initial conditions $|\Psi(0)\rangle = |1-,1-\rangle$ (solid line) and $|\Psi(0)\rangle = (|2-,0g\rangle + |0g,2-\rangle)/\sqrt{2}$. The inset shows the PQM circuit analogy, Eqs~(\ref{memristor}) and~(\ref{memristor1}). For simulations we use the parameters $g=1$, $J=10^{-2}g$, $\xi_i = 10g$, $\xi_f = 10^{3}g$, $\sigma = T/4$, and $\omega^{\rm c} = 10^4g$, and $\gamma_i^{\rm a,c} = 0$ for each cavity $i=1,2$.}
\label{fig_plas}
\end{figure}

Quantum platforms as non-equilibrium photonic systems also display hysteresis loops as a  function of the photonic number~\cite{Casteels}. However, these systems are independent of the initial condition. For instance, consider a single cavity driven by a continuous wave with frequency $\omega^{p}$ and coupling strength $I(t)$, while the atom is driven with frequency $\omega^l$ and Rabi coupling $\Omega$. In a multi-rotating frame with the atom and cavity frequencies, we get

\begin{eqnarray}
H_{\rm 1}(t) = \Delta_a \sigma^{\dagger}\sigma +  \Delta_c a^{\dagger} a +  g\left(a^{\dagger}\sigma e^{i\Delta_1 t} + a\sigma^{\dagger} e^{-i\Delta_1 t} \right) + i\Omega(\sigma^{\dagger}-\sigma)+ iI(t)\left(a^{\dagger} - a \right),
\end{eqnarray}

where $\Delta_a = \omega^{\rm a} - \omega^{\rm l}$, $\Delta_c = \omega^{\rm c}-\omega^{\rm p}$, and $\Delta_1 = \omega^{\rm p}-\omega^{\rm l}$. Photonic and atomic losses for individual cavities are described by the operator $\mathcal{D} \rho$ defined in Eq.~\eqref{Losses}. For simplicity we set $\Omega = 10^{-6}g \ll 1$, and we model $I(t) = I_0e^{-(t-T)^2/2\sigma_w^2}$ with a Gaussian profile by setting $I_0 = g$, $\sigma = T/4$, and $T = 10^3g^{-1}$. In Fig.~\ref{fig_plas} (a) we show that the system undergoes a hysteretic behavior for the resonant condition $\Delta_a = \Delta_c = \Delta_1 = 0$, but the loops remain unaffected by the initial conditions. In addition, we consider the driven non-linear Kerr model, described in Ref~\cite{Casteels}, where the system Hamiltonian in the rotating frame at the pumping frequency $\omega^{\rm p}$ is:

\begin{equation}
H_{\rm KM}(t) = -\Delta a^{\dagger}a + \frac{U}{2}(a^{\dagger})^2 a^2 + F(t)(a^{\dagger}+a),
\end{equation}

where $\Delta = \omega^{\rm p}- \omega^{\rm a}$ and $F(t) = F_0 + (t/t_s) \Delta F \theta(t_s-t)- (t- 2t_s)/t_s \Delta F \theta(t-t_s)$ represent a triangular shape, with $\Delta F$ the maximum and $\theta(t)$ the Heaviside step function. For simulating the non-linear Kerr model we use the master equation $\dot{\rho} = -i[H_{\rm KM}(t) ,\rho]+ \gamma/2(2a  \rho a^{\dagger}-\{a^{\dagger}a,\rho\})$. Once again, we observe dynamical hysteresis without a dependency on the initial conditions, see Fig.~\ref{fig_plas} (b). To illustrate our results further, we consider a pulse train for the external modulation $\xi(t)$. After each pulse, the system accumulates a phase due to the shifts in the population (see Fig.~\ref{fig_hys}(b)), and ends in a different state, which serves as the initial condition for the incoming pulse. Henceforth, we find in Fig.~\ref{fig_plas} (c) a spiral time-dependent hysteresis loops. We use the term \textit{plasticity} to denote this particular dynamical sensitiveness to initial states in our scheme. \par

Having demonstrated the notion of plasticity, a natural question arises: can our model be mathematically connected with the underlying physics of circuit elements with memory? In particular, this will be useful for physical implementations of quantum neuromorphic computing~\cite{Pfeiffer,Grollier} which simulate artificial or biological neural networks.\par

\section*{Circuit analogy}

In classical resistive memories \cite{11_memory_materials}, the current-voltage (I-V) relation $I(t)=G(x(t),V(t),t) V(t)$ holds, where $G$ is the memory conductance, and the internal state variables $x(t)$ track the past state history of the system according to the applied input signal. The inclusion of an updating function $\dot{x}(t)=f(x(t),V(t),t)$ typically yields non-linear dynamics under periodic driving and produces an I-V hysteresis loop \cite{11_memory_materials}. To make a connection between the above circuit elements and our PQM, we use the time-dependent variation of the polariton variance and the Lindblad master equation to obtain

\begin{equation}\label{output-input-equation}
\frac{dy}{d\xi} \dot{\xi} + b(\xi,\rho,t)\xi(t) = a(t,\xi,\rho,t),
\end{equation}

where $y(t) = \text{Re}\{\text{var}[N_i] \}$ accounts for the real part of the response function, $a(t,\xi,\rho,t) = \text{Re}\{-i \langle [\alpha_i,H_0] \rangle+ \text{Tr}[\alpha_i{\cal D}\rho] \}$, $b(t) = \text{Re}\{i \langle [\alpha_i,H_i] \rangle\}$ are dynamical functions that depend on the input function $\xi(t)$, the state of the system $\rho(t)$ and the Hamiltonians $H_0$ and $H_i = \sum_i \sigma_i^{\dagger}\sigma_i$. Here, $\alpha_i = (N_i-\langle N_i \rangle)^2-\langle N_i \rangle^2$ with $\langle N_i \rangle = \text{Tr}(N_i \rho)$. We note that the form of the dynamical system given in Eq.~\eqref{output-input-equation} resembles a forced RC circuit, where the following analogies can be recognized: $i(t) \rightarrow \dot{\xi}$ (current), $R \rightarrow dy/d\xi$ (resistance), $C \rightarrow b$ (capacitance), and $\varepsilon \rightarrow a$ (voltage), see Fig.~\ref{fig_plas}(d). If we define $R(\xi, \rho, t)= dy/d\xi$ as \textit{differential resistance}, we obtain the following Ohmic-like law:

\begin{eqnarray} \label{memristor}
V_R &=& R(\xi,\rho, t) \dot{\xi}, \\
\dot{\rho} &=& f(\rho, \xi, t).\label{memristor1}
\end{eqnarray}

As shown in Fig.~\ref{fig_plas} (d), the above system mimics the dynamics of a current-controlled memristive system~\cite{Pfeiffer}, where the voltage-like $V_R = \dot{y}$ varies as a function of the current $\dot{\xi}$, and the internal state evolves following a Markovian model $f(\rho, \xi, t) = -i[H(t),\rho]+{\cal D}\rho$. We remark that $R(\xi, \rho, t)= dy/d\xi$ can be positive, zero or even negative, and such behavior is connected with the initial condition and the relative time $\tau-T$. For instance, as shown in Fig.~\ref{fig_hys_open} (a) for $T= 0.95 \tau$, the variance starts from $y(t=0) = 0$ and increases as a function of the detuning $\xi$, which implies that initially $R(\xi, \rho, t)= dy/d\xi >0$. On the contrary, in Fig.~\ref{fig_hys_open} (c) for $T= 0.95 \tau$, the initial value is $y(t=0)=1$ and the variance decreases in terms of $\xi$, thus $R(\xi, \rho, t)= dy/d\xi <0$. \par

\begin{table}
\centering
\caption{Characteristics of different types of quantum memristors.}
 \label{table}
\begin{tabular}{ ||p{3cm}||p{3cm}|p{3cm}|p{4cm}|  }
 \hline
 \multicolumn{4}{|c|}{Quantum Memristors} \\
 \hline
System   & Mechanisms & Features & Implementation Status\\
 \hline
 Cavity QED polariton   & polariton exchange and external driving   & Sensitiveness to initial states, controlled memory effect without training, robust behavior against dissipative effects
.  &  Theoretical \\
 \hline
 Beam splitter with a tunable
reflectivity\cite{Raya}&  Classical feedback   &  Resistive Memory, fast update speed & Experimental \\
 \hline
 Superconducting Circuits  & Quantum feedback~\cite{Pfeiffer} or quasi-particle tunneling~\cite{Salmilehto}&  Sensitiveness to system initialization~\cite{Salmilehto}, Low temperature management, training,  &
Theoretical \\
 \hline
 Double quantum dot\cite{Ying} & Capacitive coupling & Low temperature management, small tunnel barrier & Theoretical\\
 \hline
\end{tabular}
\end{table}

As an example, let us consider two initial conditions $|\Psi(0)\rangle = |1-,1-\rangle$ (solid line with resistance $R_1$) and $|\Psi(0)\rangle = (|2-,0g\rangle + |0g,2-\rangle)/\sqrt{2}$ (dashed line with resistance $R_2$). In both cases, we have $V_R(t=0) = 0$ and $\gamma_i^{\rm a,c} = 0$ ($i=1,2$). The simulations illustrated in Fig.~\ref{fig_plas} (d) show two pinched hysteresis loops for the voltage $V_R$ as a function of the current $\dot{\xi}$, where both curves phenomenologically describe the memristive element physics. Finally, we numerically corroborate that the only difference between the two curves in Fig.~\ref{fig_plas} (d) is the sign of the resistances, \textit{i.e.} $R_1 = -R_2$, which explains the asymmetrical aspect. Finally, there have been a few suggestions of quantum memristors. In Table\ref{table} we report a comparative analysis between currently proposed implementations. The PQM presented here shares some features with superconducting circuits, in the sense that both use quasi-particle manipulation while offering remarkable tunability. 

\section*{Conclusions}

We have proposed a novel polariton-based quantum memristor (PQM), where the memristive behavior arises from the inter-cavity polariton exchange and is controlled by the time-varying atom-cavity detuning (atomic modulation). The hysteresis loop area and circulation depend on the quantum state initialization and time scales between the system and modulation. This behavior emulates a sort of ``plasticity of memristive devices''. Remarkably, this plasticity is not observed in hysteretic systems such as the single Jaynes-Cummings and non-linear Kerr model. This suggests that quasi-particle dynamics offer additional functionalities. From a technical point of view, implementing an elementary cavity QED polariton quantum memristor, whose memory content is controlled by the state initialization, adds interesting features like room temperature operation and high dynamical speed. However, perhaps the most significant aspect of PQMs, as opposed to classical memristive elements, may be the energy-efficient flow of quasi-particles that does not involve thermal dissipation due to the Joule heating. This essential feature is crucial for low-energy applications, and it makes the PQM an exciting and versatile candidate for the use of strongly-correlated light-matter systems in the field of memcomputing and (neuromorphic) quantum computation.

\bibliography{polaritons_ref_v2}

\section*{Acknowledgements}
The authors thank Peter Rabl for helpful insights and comments on the manuscript. A.N. acknowledges financial support from Universidad Mayor through the Postdoctoral Fellowship. RC acknowledges financial support from Fondecyt Iniciaci\'on No. 11180143. F. T. acknowledges financial support from Fondo Nacional de Investigaciones Cient\'ificas y Tecnol\'ogicas (FONDECYT, Chile) under grants 1211902 and Centro de Nanociencia y Nanoteccnolog\'ia CEDENNA, Financiamiento Basal para Centros Científicos y Tecnológicos de Excelencia AFB180001. M.D.  acknowledges financial support from the Department of Energy under Grant No. DE-SC0020892.

\section*{Author contributions}
A. N. and R. C. created the theoretical model. F. T. conceived the connection with memristive elements, and M. D. realized insightful comments and observations related to memristor physics. A. N. performed all numerical simulations. All authors prepared the manuscript, proofread the paper, made comments, and approved the manuscript.

\section*{Competing interests}
M.D. is the co-founder of MemComputing, Inc. (https://memcpu.com/) that is attempting to commercialize the memcomputing technology. All other authors declare no competing interests.

\end{document}